\documentstyle[12pt,epsfig]{article}
\input epsf

\begin{document}

\thispagestyle{empty}
\renewcommand{\thefootnote}{\fnsymbol{footnote}}

\begin{flushright}
{\small
SLAC--PUB--7974\\
October 1998\\}
\end{flushright}

\vspace{.8cm}

\begin{center}
{\large \bf
Identified Hadron Production and Light Quark Fragmentation in $Z^{0}$ Decays
\footnote{Work supported by Department of Energy contracts:
  DE-FG02- 91ER40676 (BU),
  DE-FG03- 91ER40618 (UCSB),
  DE-FG03- 92ER40689 (UCSC),
  DE-FG03- 93ER40788 (CSU),
  DE-FG02- 91ER40672 (Colorado),
  DE-FG02- 91ER40677 (Illinois),
  DE-AC03- 76SF00098 (LBL),
  DE-FG02- 92ER40715 (Massachusetts),
  DE-FC02- 94ER40818 (MIT),
  DE-FG03- 96ER40969 (Oregon),
  DE-AC03- 76SF00515 (SLAC),
  DE-FG05- 91ER40627 (Tennessee),
  DE-FG02- 95ER40896 (Wisconsin),
  DE-FG02- 92ER40704 (Yale);
  National Science Foundation grants:
  PHY-91- 13428 (UCSC),
  PHY-89- 21320 (Columbia),
  PHY-92- 04239 (Cincinnati),
  PHY-95- 10439 (Rutgers),
  PHY-88- 19316 (Vanderbilt),
  PHY-92- 03212 (Washington);
  The UK Particle Physics and Astronomy Research Council
  (Brunel, Oxford and RAL);
  The Istituto Nazionale di Fisica Nucleare of Italy
  (Bologna, Ferrara, Frascati, Pisa, Padova, Perugia);
  The Japan-US Cooperative Research Project on High Energy Physics
  (Nagoya, Tohoku);
  The Korea Research Foundation (Soongsil, 1997).}}

\vspace{1cm}

M. Kalelkar\\
Rutgers University, Piscataway, NJ 08855\\
\vspace{1cm}

Representing The SLD Collaboration$^{**}$\\
Stanford Linear Accelerator Center, Stanford University,
Stanford, CA  94309\\

\end{center}

\vfill

\begin{center}
{\large \bf
Abstract }
\end{center}

{\bf 
We have measured the differential cross sections for the production of
$\pi^{+}$, $K^{+}$, $K^{0}$, $K^{*0}$, $\phi$, p, $\Lambda$ and their
corresponding antiparticles in separate samples of flavor-tagged
$Z^{0}\rightarrow$ light-flavor ($u\bar{u}$, $d\bar{d}$, or $s\bar{s}$),
$Z^{0}\rightarrow c\bar{c}$ and $Z^{0}\rightarrow b\bar{b}$ events.  Clear
flavor dependences are observed, and the results are compared with the
predictions of three fragmentation models.  We have also performed a direct
measurement of $A_{s}$, the parity-violating coupling of the $Z^{0}$ to strange
quarks, by measuring the left-right-forward-backward production asymmetry in
polar angle of the tagged $s$ quark.  Our preliminary result is $A_{s}$ =
$0.82\pm 0.10(stat.)\pm 0.07(syst.)$.
}
\vfill

\begin{center} 

{\it Presented at the International Euroconference on Quantum Chromodynamics
(QCD 98), 2-8 July 1998, Montpellier, France}
\end{center}

\newpage



%
\pagestyle{plain}

\section{Introduction}

There are several phenomenological models of jet fragmentation in the process
$e^{+}e^{-} \rightarrow Z^{0} \rightarrow q\bar{q}$, followed by the radiation
of gluons and the eventual transformation of partons into primary hadrons.  The
HERWIG~\cite{herwig} model splits gluons into $q\bar{q}$ pairs, and these
quarks and antiquarks are then paired up locally to form colorless clusters
that decay into hadrons.  The JETSET~\cite{jetset} model represents the color
field between partons by a string which fragments into several pieces that
correspond to primary hadrons.  In the UCLA~\cite{ucla} model, whole events are
generated according to weights derived from phase space and Clebsch-Gordan
coefficients.

In this paper we report a measurement of the differential cross sections for
the production of $\pi^{+}$, $K^{+}$, $K^{0}$, $K^{*0}$, $\phi$, p, $\Lambda$
and their corresponding antiparticles in separate samples of flavor-tagged
$Z^{0}\rightarrow$ light-flavor ($u\bar{u}$, $d\bar{d}$, or $s\bar{s}$),
$Z^{0}\rightarrow c\bar{c}$ and $Z^{0}\rightarrow b\bar{b}$ events.  We have
used the results to test the predictions of the three fragmentation models
described above.

Measurements of the fermion production asymmetries in the process $e^{+}e^{-}
\rightarrow Z^{0} \rightarrow f\bar{f}$ provide information on the extent of
parity violation in the coupling of the $Z^{0}$ bosons to fermions of type $f$. 
The differential production cross section can be expressed in terms of $x =
\cos\theta$, where $\theta$ is the polar angle of the final state fermion $f$
with respect to the electron beam direction: $${d\sigma\over{dx}} \propto
(1-A_{e}P_{e})(1+x^{2}) + 2A_{f}(A_{e}-P_{e})x$$ where $P_{e}$ is the
longitudinal polarization of the electron beam, the positron beam is assumed
unpolarized, and the asymmetry parameters $A_{f} = 2v_{f}a_{f}/(v_{f}^{2} +
a_{f}^{2})$ are defined in terms of the vector and axial-vector couplings of
the $Z^{0}$ to fermion $f$.  The Standard Model (SM) predictions for the values
of the asymmetry parameters, assuming $\sin^{2}\theta_{w} = 0.23$, are $A_{e} =
A_{\mu} = A_{\tau} = 0.16$, $A_{u} = A_{c} = A_{t} = 0.67$, and $A_{d} = A_{s}
= A_{b} = 0.94$.  For a given final state $f\bar{f}$, if one measures the polar
angle distributions in equal luminosity samples taken with negative and
positive beam polarization, then one can derive the left-right-forward-backward
asymmetry: $$\tilde{A}^{f}_{FB} = {3\over{4}}\mid P_{e}\mid A_{f}$$ which is
insensitive to the initial state coupling.

A number of previous measurements have been made of the leptonic asymmetries
and the heavy-flavor asymmetries, but very few measurements exist for the light
quark flavors, due to the difficulty of tagging specific light flavors. We
present a direct measurement of the strange quark asymmetry parameter $A_{s}$,
in which identified strange particles are used to tag $s$ and $\bar{s}$ jets. 
The $ud$ background is suppressed by requiring a tag in both jets, and the
background is measured in the data.

In our experiment, events were produced by the SLAC Linear Collider (SLC) and
recorded in the SLC Large Detector (SLD).  The SLC delivered an electron beam
with an average polarization of 74\% and an unpolarized positron beam.  A
description of the SLD detector, trigger, track and hadronic event selection,
and Monte Carlo simulation is given in Ref.~\cite{impact}.

\section{Particle Identification}

The identification of $\pi^{\pm}$, K$^{\pm}$, p, and $\bar{\rm p}$ was achieved
by reconstructing emission angles of individual Cherenkov photons radiated by
charged particles passing through liquid and gas radiator systems of the SLD
Cherenkov Ring Imaging Detector (CRID)~\cite{crid}.  In each momentum bin,
identified $\pi$, K, and p were counted, and these were unfolded using the
inverse of an identification efficiency matrix~\cite{pavel}, and corrected for
track reconstruction efficiency.  The elements of the identification efficiency
matrix were mostly measured from data, using selected $K^{0}_{S}$, $\tau$, and
$\Lambda$ decays.  A detailed Monte Carlo simulation was used to derive the
unmeasured elements in terms of these measured ones.

Candidate $K^{0}_{S}\rightarrow\pi^{+}\pi^{-}$, $\Lambda\rightarrow{\rm
p}\pi^{-}$ and $\bar{\Lambda}\rightarrow\bar{\rm p}\pi^{+}$ decays were
selected by considering all pairs of oppositely charged tracks that were
inconsistent with originating at the interaction point and passed a set of
cuts~\cite{baird} on vertex quality and flight distance.  Backgrounds from
misidentified $\Lambda$ and $K^{0}_{S}$ decays and photon conversions were
suppressed by using kinematic cuts.

Candidate $K^{*0}\rightarrow K^{+}\pi^{-}$, $\overline{K}^{*0}\rightarrow
K^{-}\pi^{+}$ decays were selected by considering all pairs of
oppositely-charged tracks that were consistent with intersecting at the
interaction point and having one but not both tracks identified in the CRID as
a kaon~\cite{dima}.  Candidate $\phi\rightarrow K^{+}K^{-}$ decays were
similarly selected, but with both tracks required to be identified as kaons.

In each momentum bin, the number of observed $K^{0}/\overline{K}^{0}$, $\Lambda
/\bar{\Lambda}$, $K^{*0}/\overline{K}^{*0}$ and $\phi$ was determined
from a fit to the appropriate invariant mass distributions.  Finally, the
signals were corrected for reconstruction efficiencies.

\section{Production Rates}

The differential cross sections for the production of $\pi^{+}$, K$^{+}$,
K$^{0}$, K$^{*0}$, $\phi$, p, $\Lambda$ and their corresponding
antiparticles were measured as a function of the scaled momentum $x_{p} =
2p/\sqrt{s}$ of the hadron, where $p$ is its magnitude of momentum and
$\sqrt{s}$ is the $e^{+}e^{-}$ center-of-mass energy.  The SLD Vertex
Detector~\cite{vxd} was used to select subsamples of events flavor-tagged as
light ($u\bar{u}, d\bar{d}, s\bar{s}$), $c\bar{c}$, or $b\bar{b}$.  These
selections were based on impact parameters of charged tracks with respect to
the interaction point in the plane transverse to the beam.  All rates were
corrected for flavor-tagging purity and bias.

\begin{figure}[htbp]
\centering
\epsfxsize=3.4in
\leavevmode
\epsfbox{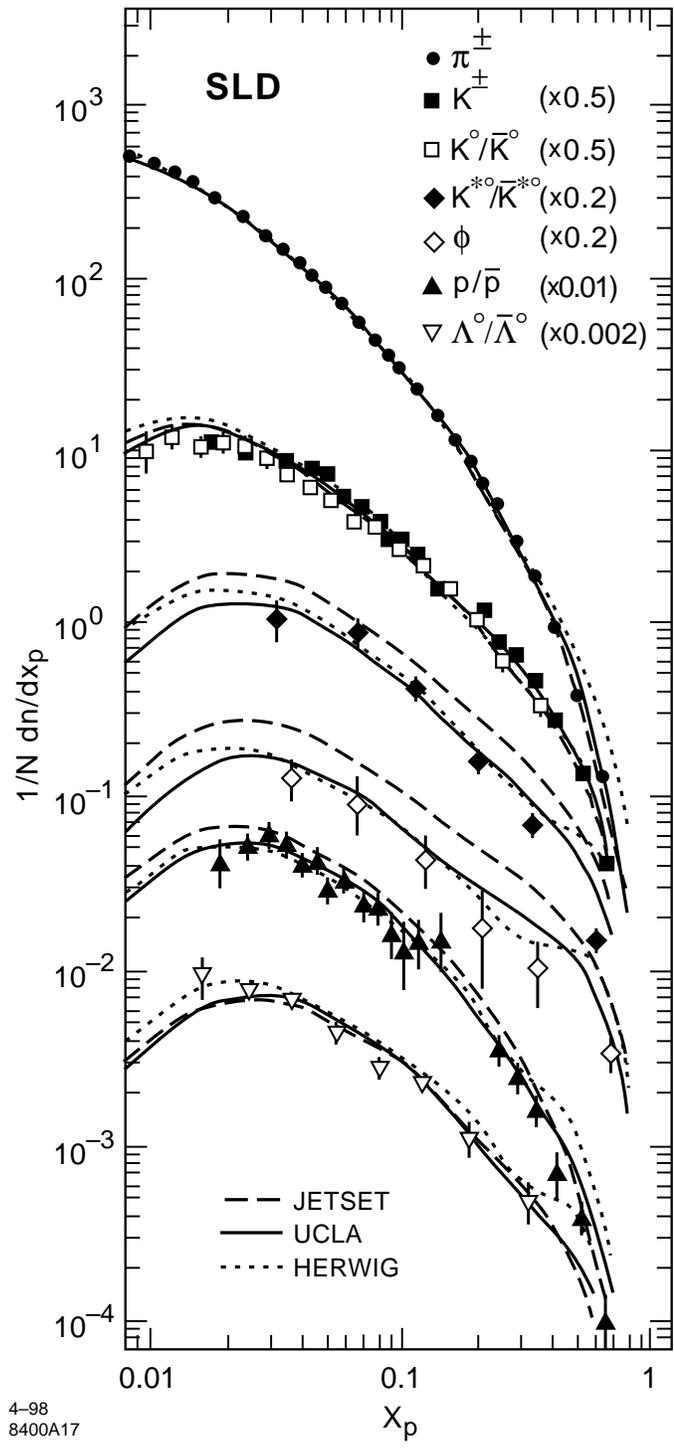}
\caption{\label{fig15}
Differential cross sections for the production of identified hadrons in the
light-flavor sample, as a function of scaled momentum.  Also shown are the
predictions of the three fragmentation models discussed in the text.}
\end{figure}

Fig.~\ref{fig15} shows the differential cross sections of the seven hadron
species in light-flavor $Z^{0}$ decays as a function of scaled momentum
$x_{p}$.  At low $x_{p}$ pions are seen to dominate over kaons by a factor of
10, and over $K^{*0}$ by a factor of 40.  Amongst the baryons at low $x_{p}$,
protons dominate over the $\Lambda^{0}$ by a factor of 3.  However, at high
$x_{p}$ the pion and kaon rates appear to be converging, as are the proton and
$\Lambda^{0}$ rates.
 
Also shown in Fig.~\ref{fig15} are the predictions of the three fragmentation
models described in the Introduction.  All the models reproduce the shape of
each differential cross section qualitatively.  The JETSET prediction for
charged pions is smaller than the data in the range $x_{p} < 0.015$, and those
for the pseudoscalar kaons are larger than the data for $0.015 < x_{p} < 0.03$;
those for the vector mesons and protons reproduce the $x_{p}$ dependence but
show a larger normalization than the data.  The HERWIG prediction for
pseudoscalar kaons is also larger than the data at low $x_{p}$ and is slightly
smaller than the data in the range $0.15 < x_{p} < 0.25$.  For all hadron
species the HERWIG prediction is larger than the data for $x_{p} > 0.4$,
showing a characteristic shoulder structure.  The UCLA predictions for the
baryons and vector mesons show a similar but less pronounced structure that is
inconsistent with the proton and $K^{*0}$ data.  Otherwise UCLA reproduces the
data except for pseudoscalar kaons in the range $0.15 < x_{p} < 0.03$.
 
\begin{figure}[htbp]
\centering
\epsfxsize=6.0in
\leavevmode
\epsfbox{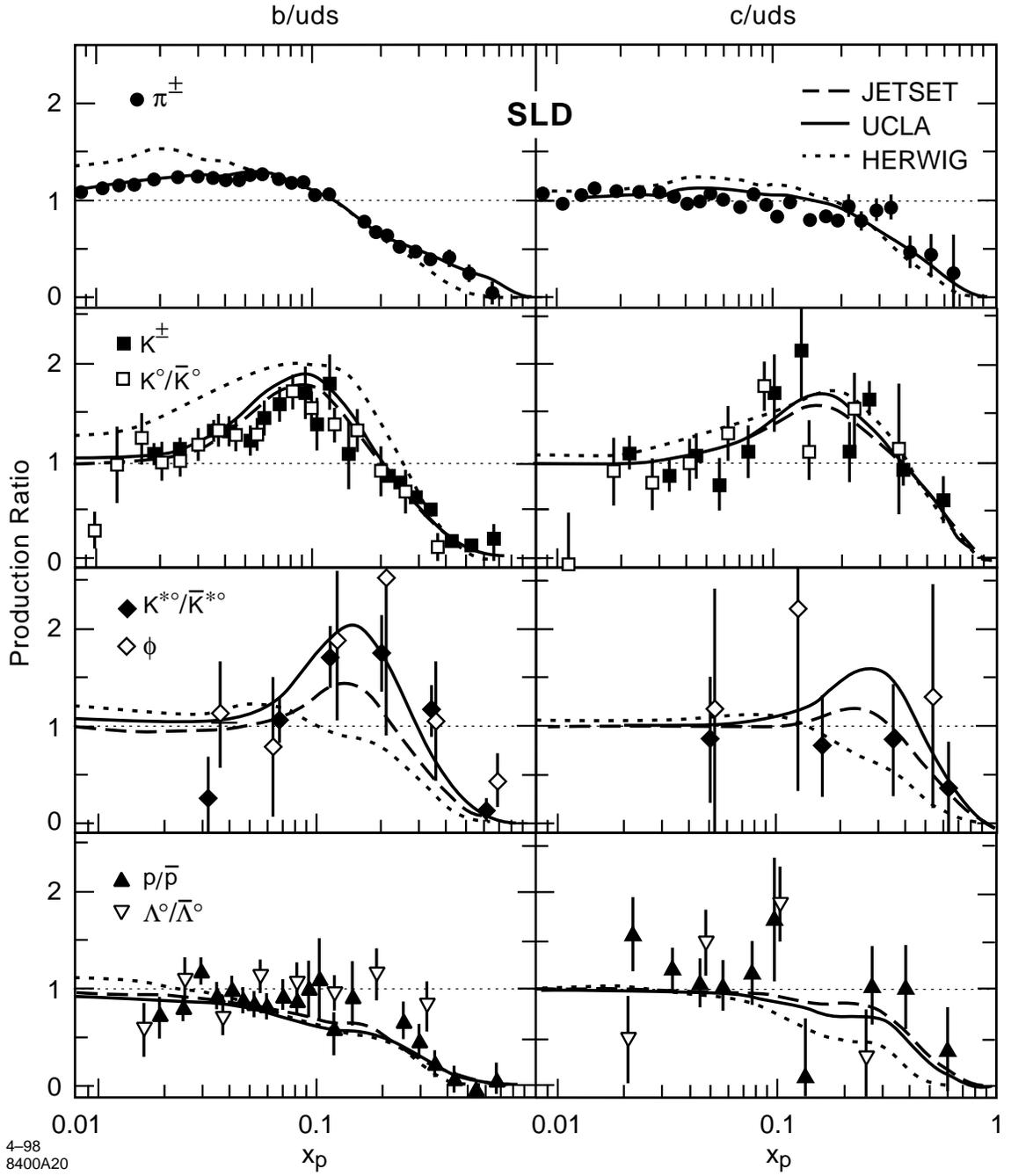}
\caption{\label{fig16}
Ratios of production of each hadron species in $b$-flavor events to that in
light-flavor events (left) and in $c$-flavor:light-flavor events (right).  Also
shown are the predictions of the three fragmentation models.}
\end{figure}

Fig.~\ref{fig16} shows the ratios of production in $b$-flavor to light flavor
events for the seven species.  
The systematic errors on the hadron
reconstruction and identification largely cancel in these ratios, and the total
errors are predominantly statistical.  There is higher production of charged
pions in $b$-flavor events than in light-flavor events at low $x_{p}$, with the
ratio rising slowly until $x_{p} = 0.06$, and falling rapidly thereafter.  The
production of both charged and neutral kaons is approximately equal in the two
samples for $x_{p} < 0.03$, but the relative production in $b$-flavor events
increases in the range $0.03 < x_{p} < 0.09$, and then decreases sharply for
higher $x_{p}$.  There is approximately equal production of baryons in the two
samples for $x_{p} < 0.15$, followed by a decline in the ratio at higher
$x_{p}$.  These features are consistent with expectations based on the known
properties of $e^{+}e^{-} \rightarrow b\bar{b}$ events, namely that a large
fraction of the event energy is carried by the leading $B$- and
$\bar{B}$-hadrons, which decay into a large number of lighter particles.

Also shown in Fig.~\ref{fig16} are the predictions of the three fragmentation
models, all of which reproduce these features qualitatively, although HERWIG
overestimates the ratio for pions in the range $x_{p} < 0.5$ and that for kaons
for $x_{p} < 0.3$.  In the right half of the figure are shown the ratios of
production in $c$-flavor to light-flavor events for the seven species. 
Features similar to those in the $b$:$uds$ comparison are observed.  There is
higher kaon production in $c$-flavor events than in light-flavor events at
$x_{p}\approx 0.1$, reflecting the tendency of $c$-jets to produce a fairly
hard charmed hadron whose decay products include a kaon carrying a large
fraction of its momentum.  Also shown are the $c$:$uds$ ratios predicted by the
fragmentation models.  All models are consistent with the data, except that
HERWIG overestimates the pion ratio for $0.03 < x_{p} < 0.15$.

\section{Strange Quark Asymmetry}

For the measurement of the strange quark asymmetry parameter $A_{s}$, the first
step was to select $s\bar{s}$ events and tag the $s$ and $\bar{s}$ jets.  Each
event was divided into two hemispheres by a plane perpendicular to the thrust
axis.  We required each hemisphere to contain at least one identified strange
($K^{\pm}$, $K^{0}_{s}$ or $\Lambda^{0}/\bar{\Lambda}^{0}$) particle, and the
strange particle of highest momentum was used to tag the strangeness of the
hemisphere.  At least one of the tagging particles was required to possess
definite strangeness (note that $K^{0}_{s}$ does not have definite
strangeness), and if both particles had definite strangeness, their strangeness
was required to be opposite.  This procedure resulted in an overall $s\bar{s}$
purity of 69\% for the selected sample.

\begin{figure}[htbp]
\centering
\epsfxsize=7.0in
\leavevmode
\epsfbox{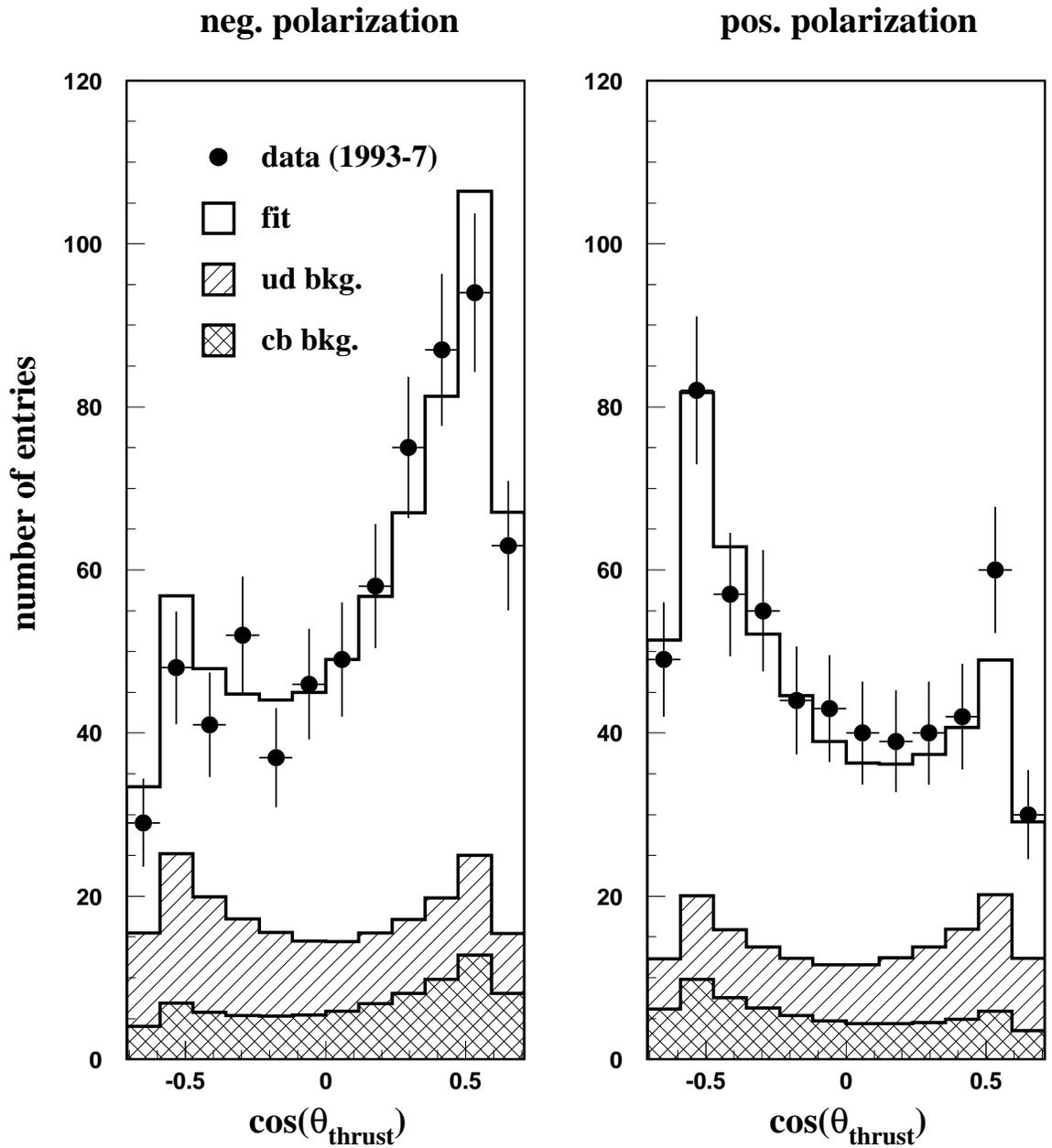}
\caption{\label{a_s}
Polar angle distributions of the thrust axis, signed with the strangeness of
the tagged strange particle, for negative (left) and positive (right) beam
polarization.  The dots show data and the histograms show our estimates of the
non-$s\bar{s}$ background and the result of the fit to the data.}
\end{figure}

The initial $s$ quark direction was approximated by the thrust axis of the
event, signed to point in the direction of negative strangeness. 
Fig.~\ref{a_s} shows the polar angle distributions of the signed thrust axis
for left handed ($P_{e}<0$) and right handed ($P_{e}>0$) electron beams.  The
expected production asymmetries, of opposite sign for the left handed and the
right handed beams, are clearly visible.
 
$A_{s}$ was extracted from these distributions by a binned maximum likelihood
fit, the result of which is shown as a histogram in the figure.  The fit
quality was good, with a $\chi^{2}$ of 12.9 for 24 bins.  
Also shown in the
figure are our estimates of the non-$s\bar{s}$ backgrounds.  The cross-hatched
histograms indicate $c\bar{c} + b\bar{b}$ backgrounds which show asymmetries of
the same sign and similar slope as the total distribution.  These backgrounds
are understood experimentally and were evaluated using a detailed Monte Carlo
simulation.  Standard systematic variations of the simulation~\cite{impact}
were considered.  The hatched histograms indicate $u\bar{u} + d\bar{d}$
backgrounds, showing asymmetries of the opposite sign and slope to the total
distribution.  This background is not well-understood experimentally, and we
are especially sensitive to both its size and slope.  Furthermore, the
analyzing power in $s\bar{s}$ events has not been measured.  Our Monte Carlo
was used to evaluate the parameters used in the fit, but was calibrated using
the data:MC ratios of the number of hemispheres with an identified $K^{+}K^{-}$
pair, the number of hemispheres with three identified kaons, and the number of
events with a $K^{\pm}$ tag of the same sign in both hemispheres.  The
uncertainties on these measured ratios were taken as systematic variations. 
Our preliminary result is $$A_{s} = 0.82\pm 0.10(stat.)\pm 0.07(syst.)$$

This result is consistent with the Standard Model expectation of 0.94 for
$A_{s}$.  Two of the LEP experiments have measured forward-backward asymmetries
from which $A_{s}$ can be derived by assuming a value for $A_{e}$.  Using
$A_{e} = 0.155$, the DELPHI~\cite{delphi} measurements translate into $A_{s} =
1.13\pm 0.30(stat.)\pm 0.11(syst.)$ and $A_{d,s} = 0.96\pm 0.27(stat.)\pm
0.46(syst.)$.  The OPAL~\cite{opal} measurements yield $A_{d,s} = 0.58\pm
0.30(stat.)\pm 0.09(syst.)$.  Our measurement is consistent with these and
represents a substantial improvement in precision.

\section{Conclusions}

We have measured the differential cross sections as a function of scaled
momentum $x_{p}$ for the production of $\pi^{+}$, K$^{+}$, K$^{0}$, K$^{*0}$,
$\phi$, p, $\Lambda$ and their corresponding antiparticles separately in
light-flavor, $c\bar{c}$ and $b\bar{b}$ jets from $Z^{0}$ decays.  Significant
differences between flavors were found, consistent with expectations based on
the known properties of $B$ and $D$ hadron production and decay.  Our data were
used to test the predictions of three fragmentation models with default
parameters.  In most cases these simulations reproduced the data to within a
few percent.  Each model, however, did disagree with some features of the data
in isolated regions of $x_{p}$.

We have also performed a measurement of $A_{s}$, the parity-violating coupling
of the $Z^{0}$ to strange quarks, obtained directly from the
left-right-forward-backward production asymmetry in polar angle of the tagged
$s$ quark.  Our preliminary result is $A_{s}$ = $0.82\pm 0.10(stat.)\pm
0.07(syst.)$, which is consistent with the Standard Model expectation.  It is
also consistent with previous measurements of $A_{s}$, but with significantly
smaller uncertainties.

\newpage
\section*{$^{**}$List of Authors} 
%
%
%
\begin{center}
\def\iADEL{$^{(1)}$}
\def\iAOMORI{$^{(2)}$}
\def\iBOLO{$^{(3)}$}
\def\iBRUN{$^{(4)}$}
\def\iBU{$^{(5)}$}
\def\iCINC{$^{(6)}$}
\def\iCOLO{$^{(7)}$}
\def\iCOLU{$^{(8)}$}
\def\iCSU{$^{(9)}$}
\def\iFERR{$^{(10)}$}
\def\iFRAS{$^{(11)}$}
\def\iILLI{$^{(12)}$}
\def\iLBL{$^{(13)}$}
\def\iLTU{$^{(14)}$}
\def\iMASS{$^{(15)}$}
\def\iMISSI{$^{(16)}$}
\def\iMIT{$^{(17)}$}
\def\iMOSCOW{$^{(18)}$}
\def\iNAGO{$^{(19)}$}
\def\iOREG{$^{(20)}$}
\def\iOXF{$^{(21)}$}
\def\iPADO{$^{(22)}$}
\def\iPERU{$^{(23)}$}
\def\iPISA{$^{(24)}$}
\def\iRAL{$^{(25)}$}
\def\iRUTG{$^{(26)}$}
\def\iSLAC{$^{(27)}$}
\def\iSOGA{$^{(28)}$}
\def\iSOONG{$^{(29)}$}
\def\iTENN{$^{(30)}$}
\def\iTOHO{$^{(31)}$}
\def\iUCSB{$^{(32)}$}
\def\iUCSC{$^{(33)}$}
\def\iVAND{$^{(34)}$}
\def\iWASH{$^{(35)}$}
\def\iWISC{$^{(36)}$}
\def\iYALE{$^{(37)}$}

  \baselineskip=.75\baselineskip  
\mbox{K. Abe\unskip,\iAOMORI}
\mbox{K.  Abe\unskip,\iNAGO}
\mbox{T. Abe\unskip,\iSLAC}
\mbox{I.Adam\unskip,\iSLAC}
\mbox{T.  Akagi\unskip,\iSLAC}
\mbox{N. J. Allen\unskip,\iBRUN}
\mbox{A. Arodzero\unskip,\iOREG}
\mbox{W.W. Ash\unskip,\iSLAC}
\mbox{D. Aston\unskip,\iSLAC}
\mbox{K.G. Baird\unskip,\iMASS}
\mbox{C. Baltay\unskip,\iYALE}
\mbox{H.R. Band\unskip,\iWISC}
\mbox{M.B. Barakat\unskip,\iLTU}
\mbox{O. Bardon\unskip,\iMIT}
\mbox{T.L. Barklow\unskip,\iSLAC}
\mbox{J.M. Bauer\unskip,\iMISSI}
\mbox{G. Bellodi\unskip,\iOXF}
\mbox{R. Ben-David\unskip,\iYALE}
\mbox{A.C. Benvenuti\unskip,\iBOLO}
\mbox{G.M. Bilei\unskip,\iPERU}
\mbox{D. Bisello\unskip,\iPADO}
\mbox{G. Blaylock\unskip,\iMASS}
\mbox{J.R. Bogart\unskip,\iSLAC}
\mbox{B. Bolen\unskip,\iMISSI}
\mbox{G.R. Bower\unskip,\iSLAC}
\mbox{J. E. Brau\unskip,\iOREG}
\mbox{M. Breidenbach\unskip,\iSLAC}
\mbox{W.M. Bugg\unskip,\iTENN}
\mbox{D. Burke\unskip,\iSLAC}
\mbox{T.H. Burnett\unskip,\iWASH}
\mbox{P.N. Burrows\unskip,\iOXF}
\mbox{A. Calcaterra\unskip,\iFRAS}
\mbox{D.O. Caldwell\unskip,\iUCSB}
\mbox{D. Calloway\unskip,\iSLAC}
\mbox{B. Camanzi\unskip,\iFERR}
\mbox{M. Carpinelli\unskip,\iPISA}
\mbox{R. Cassell\unskip,\iSLAC}
\mbox{R. Castaldi\unskip,\iPISA}
\mbox{A. Castro\unskip,\iPADO}
\mbox{M. Cavalli-Sforza\unskip,\iUCSC}
\mbox{A. Chou\unskip,\iSLAC}
\mbox{E. Church\unskip,\iWASH}
\mbox{H.O. Cohn\unskip,\iTENN}
\mbox{J.A. Coller\unskip,\iBU}
\mbox{M.R. Convery\unskip,\iSLAC}
\mbox{V. Cook\unskip,\iWASH}
\mbox{R. Cotton\unskip,\iBRUN}
\mbox{R.F. Cowan\unskip,\iMIT}
\mbox{D.G. Coyne\unskip,\iUCSC}
\mbox{G. Crawford\unskip,\iSLAC}
\mbox{C.J.S. Damerell\unskip,\iRAL}
\mbox{M. N. Danielson\unskip,\iCOLO}
\mbox{M. Daoudi\unskip,\iSLAC}
\mbox{N. de Groot\unskip,\iSLAC}
\mbox{R. Dell'Orso\unskip,\iPERU}
\mbox{P.J. Dervan\unskip,\iBRUN}
\mbox{R. de Sangro\unskip,\iFRAS}
\mbox{M. Dima\unskip,\iCSU}
\mbox{A. D'Oliveira\unskip,\iCINC}
\mbox{D.N. Dong\unskip,\iMIT}
\mbox{P.Y.C. Du\unskip,\iTENN}
\mbox{R. Dubois\unskip,\iSLAC}
\mbox{B.I. Eisenstein\unskip,\iILLI}
\mbox{V. Eschenburg\unskip,\iMISSI}
\mbox{E. Etzion\unskip,\iWISC}
\mbox{S. Fahey\unskip,\iCOLO}
\mbox{D. Falciai\unskip,\iFRAS}
\mbox{C. Fan\unskip,\iCOLO}
\mbox{J.P. Fernandez\unskip,\iUCSC}
\mbox{M.J. Fero\unskip,\iMIT}
\mbox{K.Flood\unskip,\iMASS}
\mbox{R. Frey\unskip,\iOREG}
\mbox{T. Gillman\unskip,\iRAL}
\mbox{G. Gladding\unskip,\iILLI}
\mbox{S. Gonzalez\unskip,\iMIT}
\mbox{E.L. Hart\unskip,\iTENN}
\mbox{J.L. Harton\unskip,\iCSU}
\mbox{A. Hasan\unskip,\iBRUN}
\mbox{K. Hasuko\unskip,\iTOHO}
\mbox{S. J. Hedges\unskip,\iBU}
\mbox{S.S. Hertzbach\unskip,\iMASS}
\mbox{M.D. Hildreth\unskip,\iSLAC}
\mbox{J. Huber\unskip,\iOREG}
\mbox{M.E. Huffer\unskip,\iSLAC}
\mbox{E.W. Hughes\unskip,\iSLAC}
\mbox{X.Huynh\unskip,\iSLAC}
\mbox{H. Hwang\unskip,\iOREG}
\mbox{M. Iwasaki\unskip,\iOREG}
\mbox{D. J. Jackson\unskip,\iRAL}
\mbox{P. Jacques\unskip,\iRUTG}
\mbox{J.A. Jaros\unskip,\iSLAC}
\mbox{Z.Y. Jiang\unskip,\iSLAC}
\mbox{A.S. Johnson\unskip,\iSLAC}
\mbox{J.R. Johnson\unskip,\iWISC}
\mbox{R.A. Johnson\unskip,\iCINC}
\mbox{T. Junk\unskip,\iSLAC}
\mbox{R. Kajikawa\unskip,\iNAGO}
\mbox{M. Kalelkar\unskip,\iRUTG}
\mbox{Y. Kamyshkov\unskip,\iTENN}
\mbox{H.J. Kang\unskip,\iRUTG}
\mbox{I. Karliner\unskip,\iILLI}
\mbox{H. Kawahara\unskip,\iSLAC}
\mbox{Y. D. Kim\unskip,\iSOGA}
\mbox{R. King\unskip,\iSLAC}
\mbox{M.E. King\unskip,\iSLAC}
\mbox{R.R. Kofler\unskip,\iMASS}
\mbox{N.M. Krishna\unskip,\iCOLO}
\mbox{R.S. Kroeger\unskip,\iMISSI}
\mbox{M. Langston\unskip,\iOREG}
\mbox{A. Lath\unskip,\iMIT}
\mbox{D.W.G. Leith\unskip,\iSLAC}
\mbox{V. Lia\unskip,\iMIT}
\mbox{C.-J. S. Lin\unskip,\iSLAC}
\mbox{X. Liu\unskip,\iUCSC}
\mbox{M.X. Liu\unskip,\iYALE}
\mbox{M. Loreti\unskip,\iPADO}
\mbox{A. Lu\unskip,\iUCSB}
\mbox{H.L. Lynch\unskip,\iSLAC}
\mbox{J. Ma\unskip,\iWASH}
\mbox{G. Mancinelli\unskip,\iRUTG}
\mbox{S. Manly\unskip,\iYALE}
\mbox{G. Mantovani\unskip,\iPERU}
\mbox{T.W. Markiewicz\unskip,\iSLAC}
\mbox{T. Maruyama\unskip,\iSLAC}
\mbox{H. Masuda\unskip,\iSLAC}
\mbox{E. Mazzucato\unskip,\iFERR}
\mbox{A.K. McKemey\unskip,\iBRUN}
\mbox{B.T. Meadows\unskip,\iCINC}
\mbox{G. Menegatti\unskip,\iFERR}
\mbox{R. Messner\unskip,\iSLAC}
\mbox{P.M. Mockett\unskip,\iWASH}
\mbox{K.C. Moffeit\unskip,\iSLAC}
\mbox{T.B. Moore\unskip,\iYALE}
\mbox{M.Morii\unskip,\iSLAC}
\mbox{D. Muller\unskip,\iSLAC}
\mbox{V.Murzin\unskip,\iMOSCOW}
\mbox{T. Nagamine\unskip,\iTOHO}
\mbox{S. Narita\unskip,\iTOHO}
\mbox{U. Nauenberg\unskip,\iCOLO}
\mbox{H. Neal\unskip,\iSLAC}
\mbox{M. Nussbaum\unskip,\iCINC}
\mbox{N.Oishi\unskip,\iNAGO}
\mbox{D. Onoprienko\unskip,\iTENN}
\mbox{L.S. Osborne\unskip,\iMIT}
\mbox{R.S. Panvini\unskip,\iVAND}
\mbox{H. Park\unskip,\iOREG}
\mbox{C. H. Park\unskip,\iSOONG}
\mbox{T.J. Pavel\unskip,\iSLAC}
\mbox{I. Peruzzi\unskip,\iFRAS}
\mbox{M. Piccolo\unskip,\iFRAS}
\mbox{L. Piemontese\unskip,\iFERR}
\mbox{E. Pieroni\unskip,\iPISA}
\mbox{K.T. Pitts\unskip,\iOREG}
\mbox{R.J. Plano\unskip,\iRUTG}
\mbox{R. Prepost\unskip,\iWISC}
\mbox{C.Y. Prescott\unskip,\iSLAC}
\mbox{G.D. Punkar\unskip,\iSLAC}
\mbox{J. Quigley\unskip,\iMIT}
\mbox{B.N. Ratcliff\unskip,\iSLAC}
\mbox{T.W. Reeves\unskip,\iVAND}
\mbox{J. Reidy\unskip,\iMISSI}
\mbox{P.L. Reinertsen\unskip,\iUCSC}
\mbox{P.E. Rensing\unskip,\iSLAC}
\mbox{L.S. Rochester\unskip,\iSLAC}
\mbox{P.C. Rowson\unskip,\iCOLU}
\mbox{J.J. Russell\unskip,\iSLAC}
\mbox{O.H. Saxton\unskip,\iSLAC}
\mbox{T. Schalk\unskip,\iUCSC}
\mbox{R.H. Schindler\unskip,\iSLAC}
\mbox{B.A. Schumm\unskip,\iUCSC}
\mbox{J. Schwiening\unskip,\iSLAC}
\mbox{S. Sen\unskip,\iYALE}
\mbox{V.V. Serbo\unskip,\iWISC}
\mbox{M.H. Shaevitz\unskip,\iCOLU}
\mbox{J.T. Shank\unskip,\iBU}
\mbox{G. Shapiro\unskip,\iLBL}
\mbox{D.J. Sherden\unskip,\iSLAC}
\mbox{K. D. Shmakov\unskip,\iTENN}
\mbox{C. Simopoulos\unskip,\iSLAC}
\mbox{N.B. Sinev\unskip,\iOREG}
\mbox{S.R. Smith\unskip,\iSLAC}
\mbox{M. B. Smy\unskip,\iCSU}
\mbox{J.A. Snyder\unskip,\iYALE}
\mbox{H. Staengle\unskip,\iCSU}
\mbox{A. Stahl\unskip,\iSLAC}
\mbox{P. Stamer\unskip,\iRUTG}
\mbox{R. Steiner\unskip,\iADEL}
\mbox{H. Steiner\unskip,\iLBL}
\mbox{M.G. Strauss\unskip,\iMASS}
\mbox{D. Su\unskip,\iSLAC}
\mbox{F. Suekane\unskip,\iTOHO}
\mbox{A. Sugiyama\unskip,\iNAGO}
\mbox{S. Suzuki\unskip,\iNAGO}
\mbox{M. Swartz\unskip,\iSLAC}
\mbox{A. Szumilo\unskip,\iWASH}
\mbox{T. Takahashi\unskip,\iSLAC}
\mbox{F.E. Taylor\unskip,\iMIT}
\mbox{J. Thom\unskip,\iSLAC}
\mbox{E. Torrence\unskip,\iMIT}
\mbox{N. K. Toumbas\unskip,\iSLAC}
\mbox{A.I. Trandafir\unskip,\iMASS}
\mbox{J.D. Turk\unskip,\iYALE}
\mbox{T. Usher\unskip,\iSLAC}
\mbox{C. Vannini\unskip,\iPISA}
\mbox{J. Va'vra\unskip,\iSLAC}
\mbox{E. Vella\unskip,\iSLAC}
\mbox{J.P. Venuti\unskip,\iVAND}
\mbox{R. Verdier\unskip,\iMIT}
\mbox{P.G. Verdini\unskip,\iPISA}
\mbox{S.R. Wagner\unskip,\iSLAC}
\mbox{D. L. Wagner\unskip,\iCOLO}
\mbox{A.P. Waite\unskip,\iSLAC}
\mbox{Walston, S.\unskip,\iOREG}
\mbox{J.Wang\unskip,\iSLAC}
\mbox{C. Ward\unskip,\iBRUN}
\mbox{S.J. Watts\unskip,\iBRUN}
\mbox{A.W. Weidemann\unskip,\iTENN}
\mbox{E. R. Weiss\unskip,\iWASH}
\mbox{J.S. Whitaker\unskip,\iBU}
\mbox{S.L. White\unskip,\iTENN}
\mbox{F.J. Wickens\unskip,\iRAL}
\mbox{B. Williams\unskip,\iCOLO}
\mbox{D.C. Williams\unskip,\iMIT}
\mbox{S.H. Williams\unskip,\iSLAC}
\mbox{S. Willocq\unskip,\iSLAC}
\mbox{R.J. Wilson\unskip,\iCSU}
\mbox{W.J. Wisniewski\unskip,\iSLAC}
\mbox{J. L. Wittlin\unskip,\iMASS}
\mbox{M. Woods\unskip,\iSLAC}
\mbox{G.B. Word\unskip,\iVAND}
\mbox{T.R. Wright\unskip,\iWISC}
\mbox{J. Wyss\unskip,\iPADO}
\mbox{R.K. Yamamoto\unskip,\iMIT}
\mbox{J.M. Yamartino\unskip,\iMIT}
\mbox{X. Yang\unskip,\iOREG}
\mbox{J. Yashima\unskip,\iTOHO}
\mbox{S.J. Yellin\unskip,\iUCSB}
\mbox{C.C. Young\unskip,\iSLAC}
\mbox{H. Yuta\unskip,\iAOMORI}
\mbox{G. Zapalac\unskip,\iWISC}
\mbox{R.W. Zdarko\unskip,\iSLAC}
\mbox{J. Zhou\unskip.\iOREG}

\it
  \vskip \baselineskip                   
  \centerline{(The SLD Collaboration)}   
  \vskip \baselineskip        
  \baselineskip=.75\baselineskip   
\iADEL
  Adelphi University,
  South Avenue-   Garden City,NY 11530, \break
\iAOMORI
  Aomori University,
  2-3-1 Kohata, Aomori City, 030 Japan, \break
\iBOLO
  INFN Sezione di Bologna,
  Via Irnerio 46    I-40126 Bologna  (Italy), \break
\iBRUN
  Brunel University,
  Uxbridge, Middlesex - UB8 3PH United Kingdom, \break
\iBU
  Boston University,
  590 Commonwealth Ave. - Boston,MA 02215, \break
\iCINC
  University of Cincinnati,
  Cincinnati,OH 45221, \break
\iCOLO
  University of Colorado,
  Campus Box 390 - Boulder,CO 80309, \break
\iCOLU
  Columbia University,
  Nevis Laboratories  P.O.Box 137 - Irvington,NY 10533, \break
\iCSU
  Colorado State University,
  Ft. Collins,CO 80523, \break
\iFERR
  INFN Sezione di Ferrara,
  Via Paradiso,12 - I-44100 Ferrara (Italy), \break
\iFRAS
  Lab. Nazionali di Frascati,
  Casella Postale 13   I-00044 Frascati (Italy), \break
\iILLI
  University of Illinois,
  1110 West Green St.  Urbana,IL 61801, \break
\iLBL
  Lawrence Berkeley Laboratory,
  Dept.of Physics 50B-5211 University of California-  Berkeley,CA 94720, \break
\iLTU
  Louisiana Technical University,
  , \break
\iMASS
  University of Massachusetts,
  Amherst,MA 01003, \break
\iMISSI
  University of Mississippi,
  University,MS 38677, \break
\iMIT
  Massachusetts Institute of Technology,
  77 Massachussetts Avenue  Cambridge,MA 02139, \break
\iMOSCOW
  Moscow State University,
  Institute of Nuclear Physics  119899 Moscow  Russia, \break
\iNAGO
  Nagoya University,
  Nagoya 464 Japan, \break
\iOREG
  University of Oregon,
  Department of Physics  Eugene,OR 97403, \break
\iOXF
  Oxford University,
  Oxford, OX1 3RH, United Kingdom, \break
\iPADO
  Universita di Padova,
  Via F. Marzolo,8   I-35100 Padova (Italy), \break
\iPERU
  Universita di Perugia, Sezione INFN,
  Via A. Pascoli  I-06100 Perugia (Italy), \break
\iPISA
  INFN, Sezione di Pisa,
  Via Livornese,582/AS  Piero a Grado  I-56010 Pisa (Italy), \break
\iRAL
  Rutherford Appleton Laboratory,
  Chiton,Didcot - Oxon OX11 0QX United Kingdom, \break
\iRUTG
  Rutgers University,
  Serin Physics Labs  Piscataway,NJ 08855-0849, \break
\iSLAC
  Stanford Linear Accelerator Center,
  2575 Sand Hill Road  Menlo Park,CA 94025, \break
\iSOGA
  Sogang University,
  Ricci Hall  Seoul, Korea, \break
\iSOONG
  Soongsil University,
  Dongjakgu Sangdo 5 dong 1-1    Seoul, Korea 156-743, \break
\iTENN
  University of Tennessee,
  401 A.H. Nielsen Physics Blg.  -  Knoxville,Tennessee 37996-1200, \break
\iTOHO
  Tohoku University,
  Bubble Chamber Lab. - Aramaki - Sendai 980 (Japan), \break
\iUCSB
  U.C. Santa Barbara,
  3019 Broida Hall  Santa Barbara,CA 93106, \break
\iUCSC
  U.C. Santa Cruz,
  Santa Cruz,CA 95064, \break
\iVAND
  Vanderbilt University,
  Stevenson Center,Room 5333  P.O.Box 1807,Station B  Nashville,TN 37235,
\break
\iWASH
  University of Washington,
  Seattle,WA 98105, \break
\iWISC
  University of Wisconsin,
  1150 University Avenue  Madison,WS 53706, \break
\iYALE
  Yale University,
  5th Floor Gibbs Lab. - P.O.Box 208121 - New Haven,CT 06520-8121. \break
\rm
%

\end{center}

\end{document}